# AN APPROACH TO UNIFYING CLASSIC AND QUANTUM ELECTRODYNAMICS

MASSIMO AUCI
L.S.S. "G. Bruno", via Marinuzzi 1, 10156, Torino, Italy

GUIDO DEMATTEIS
I.T.I.S. "J.C. Maxwell", via XXV Aprile 141, 10042, Nichelino, Italy

The foundations of QM can find a consistent and exhaustive explanation in a new theoretical context. The Bridge Theory (BT) allows us to justify both classic and quantum electromagnetic phenomenology by using classical concepts. In this paper we review the bases of the BT originating from the role that the transverse component of the Poynting vector plays in localising energy in the neighbourhood of an electromagnetic source and we analyse the quantum implications.

## 1. Introduction

Since of the birth of QM, the idea of a hypothetical unified "Quantum-Wave" theory of light, introduced by A. Einstein in its work on the nature of light[(*)], derives from the inadequacy of both quantum and classical em theories in describing coherently the whole of phenomena associated with light. In a different context, a similar difficulty in the deep comprehension of some aspects of quantum (and classical) theory, has been expressed by R.P. Feynman in his book [1] on QED. Feynman, considering the Sommerfeld's fine structure constant

$$\alpha = e^2/\hbar c \approx 1/137,$$

writes that it seems to have been written by God's hand and that we do not know how he may have moved his pencil in deriving it.

The physical role of the fine structure constant and its theoretical connections with the most important electromagnetic (EM) constants, put in evidence how Feynman's sentence contains the same idea of Einstein's. In fact, "God's hand" should be obviously interpreted as a unique universal drawing of nature describing an (unknown) theory able to justify all the microscopic and macroscopic phenomena associated with light, matter and the electromagnetic field.

---

[(*)] A. Einstein, Uber die Entwicklung unserer Anschauungen uber das Wesen und die Konstitution der Strahlung, pp. 816-825.

Following Einstein and Feynman's idea, a new model of the dynamical behaviour of a pair of charges during their em interaction has been proposed in ref. [2-3-4]. The model allows us to propose a derivation of the value of the coupling constant $\alpha$. Planck's constant $\hbar$ as well as quantum energy and momentum exchanged during the pair interaction follow in the same phenomenological context and a coherent derivation for an uncertainty principle can also be obtained.

We think that the physical ideas contained in the model [2-3-4], if theoretically organised in a wider context, should permit the construction of a "bridge" between the phenomenology described by QM and that described by the classical em theory. In this sense the "Bridge Theory" (BT) is able to unify wave and quantum em theory, yielding a continuity between wave and quantum em phenomenology.

In this paper, we review the derivation of the BT putting a special enphasis on its classical, but non-standard, foundations. New results concerning classical and quantum em phenomenology are given.

**2. The Basic Ideas**

The theory develops from the consideration that the effective spatial symmetry that characterises an em wave depends on the nature of the source. If we consider an ideal point-like source of em waves, the propagation occurs radially with spherical wave fronts, so that also the Poynting vector (PV) will be radial. We shall call "ideal" such a source. On the other hand, in nature one never deals with "ideal" sources. The simplest source that can be produced is at least endowed with a dipole moment, usually is not point-like. Therefore the propagation will not occur by a spherical wave front and the PV will not be radial everywhere. Hence, the PV will have a transverse component different from zero. We shall call "real" such a source.

It is usual to assume that at a distance from the source much greater than the emission wavelength, the wave has total spherical symmetry. At short distances, however, this assumption introduces non-negligible approximations from the energy point of view. In fact, the theory is based on the physical effects produced by the lack of spherical symmetry in the wave emission of a "real" source. In this case, the non-zero transverse component of the PV reduces the radial emission of energy and, consequently, localises energy around the source.

**3. Electromagnetic Sources**

In order to introduce the BT, we begin by defining as "ideal" (IS) a point-like em source in which the PV **S** is radial and as "real" (RS) any source in which the PV is not radial.

As we shall prove later, the mass of the electromagnetically interacting particles is irrelevant from the point of view of the source emission, hence the charged particles will be considered as massless during their mutual em interaction.

A dipole produced by two interacting particles is a RS; consequently, the PV, perpendicular to the wave front, can be broken into a radial and transverse component.

Since in a RS the relative motion of the particles during their interaction produces a time-varying direction of the dipole moment, it is necessary to define a third kind of source that we shall call RS with time-dependent polarisation (TVP).

For both an ideal and a real source, the energy irradiated at a given frequency per unit time is the same at very large distances, but at short distances, the presence of the transverse component of the PV causes the radial component to be smaller than that of an IS. This means that at short distances a RS irradiates less energy per unit time than an IS and a certain amount of energy is localised in proximity of the source. Therefore, a RS will not instantaneously emit all its energy like an IS does, but it will require "some time" to do it.

Let us consider an observer placed at a point P of an ideal expanding spherical surface $\Sigma(t)$ inside which a RS with fixed polarization is placed. The energy per unit time pertaining to an infinitesimal portion of the expanding surface is equal to the flux of the PV **S'** through the surface $d\mathbf{a}$. So the energy observed in P during the time interval $\tau$ is given by:

$$\delta d\varepsilon_{rad} = \int_0^\tau (\mathbf{S'} \cdot d\mathbf{a}) dt = \int_0^\tau S'_r \, da \, dt \, , \qquad (1)$$

where $S'_r$ is the radial component of the PV of the real source.

Consider now wath appens from the point of view of an hypothetical observer at P. Such an observer sees the emission of the source just along the direction characterised by the angles $\theta$ and $\varphi$, but has no information about the emission along directions different from his own. He is forced to assume that the source emits in all the other directions as it does in the direction along which he observes the source, i.e. that the source has spherical symmetry as an IS. This observer will rewrite eq. (1) as

$$\delta d\varepsilon_{rad} = \int_0^\tau S \, da \, dt \, , \qquad (2)$$

where $S \equiv S_r$ as for an IS.

By eq. (2), the total energy observed in P (i.e. along the direction $(\theta, \varphi)$) during the time interval $\tau$ is

$$\delta \varepsilon_{rad} = \int_0^\tau \left( \oint_{\Sigma(t)} S \, da \right) dt = \frac{1}{c} \int_{V_{\delta\Sigma}} S \, d^3x \qquad (3)$$

where $S/c$ is the energy density inside the spherical crown of volume $V_{\delta\Sigma}$ associated with the energy produced and emitted along the direction of P.

Usually, one deals with a RS with a TVP, but, in this case, it is sufficient to treat this kind of source as one with fixed polarization.

Each observer, independently of the polarisation, can obtain the total energy produced and emitted by a RS by integrating over all the energies measured by the observers placed in each point P of the surface $\Sigma$. The irradiated energy is then given by

$$\varepsilon_{rad} = \int_{P_\Sigma} \delta \varepsilon_{rad} \, . \qquad (4)$$

Eq. (4), being independent on the dipole polarisation, is also true for a RS with a TVP.

During the arbitrary time $\tau$, each observer placed in whatever point P of space and measuring the radial emission for a RS with TVP sees that the dipole moment changes direction. The effect of this is the variation of the energy emitted per unit time along the direction of observation. During the same time, all observers measure statistically the same amount of energy.

Once the energy is known, each observer can assume that the time-variation of the radial emission, that we define as REM (Radial Energy Modulation), is equivalent to that he would observe for a source with FP, moving during the same arbitrary time around the virtual origin of the source, along a same (chaotic) path. Therefore, the two kinds of RS can be considered as equivalent as far as their emissivity is concerned. In this sense, we shall call a RS with fixed polarization "homologue" to a RS with TVP.

Now, we can reasonably assume that, during the arbitrary time interval $\tau$, the energy produced by two homologue sources is emitted in such a way that the energy emitted per unit time by a RS with TVP across an infinitesimal portion of surface $da$ equals the mean energy emitted by a RS with fixed polarization.

Accordingly, we write:

$$\int_0^\tau (\mathbf{S'}\cdot d\mathbf{a})_{TVP}\, dt \equiv \int_0^\tau \overline{(\mathbf{S'}\cdot d\mathbf{a})}_{FP}\, dt \quad ,$$

where the observer in P measures the left-hand side and assumes that it can be written in the way of the right-hand side.

To calculate $\delta\varepsilon_{rad}$ as given by eq. (3), we must integrate the right-hand side of the previous equivalence over the surface $\Sigma(t)$ before integrating it with respect to time, obtaining

$$\delta\varepsilon_{rad} = \int_0^\tau \left( \oint_{\Sigma(t)} \overline{(\mathbf{S'}\cdot d\mathbf{a})}_{FP} \right) dt = \frac{1}{c}\int_{V_{\delta\Sigma}} \overline{S}\, d^3x \,. \tag{5}$$

Equation (5) allows us to treat a RS with TVP as a source with FP. In the same equation, the source is assumed to be ideal as an effect of the independence of the mean radial component of the PV with respect to the angular direction along which the observer is forced to measure the emission during the interacting time $\tau$.

**4. Energy Localization**

If no particle comes in or out the surface $\Sigma(t)$ placed around a source, and if every point of the surface moves from the source with a uniform radial motion we can write, by Poynting's theorem [2-5], the local (i.e. referring to the direction $(\theta,\varphi)$) conservation law for the total energy $\delta\varepsilon(t)_{tot}$ inside the surface $\Sigma(t)$:

$$\delta\dot\varepsilon(t)_{tot} = \delta\dot\varepsilon(t)_{mec} + \delta\dot\varepsilon(t)_{field} = -\oint_{\Sigma(t)} S\, da \,. \tag{6}$$

If the mechanical energy which supplies the em source is provided from outside, so that $\delta\dot\varepsilon_{mec} = 0$, the total energy variation inside the spherical crown of volume $V_{\delta\Sigma}$ will depend only on the em field, thus $\delta\dot\varepsilon(t)_{tot} = \delta\dot\varepsilon(t)_{field}$. By integrating eq.(6) over the arbitrary time interval [0, $\tau$], we obtain (see eq. (3)):

$$\delta\varepsilon(\tau)_{tot} = \delta\varepsilon(0)_{tot} - \delta\varepsilon_{rad} \quad ,$$

where

$$\delta\varepsilon(0)_{tot} = \int_{V_{\delta\Sigma}} \frac{E^2+B^2}{8\pi}\, d^3x + \kappa_{mec}$$

is the local energy at time t = 0.

Hence, by eq. (3), we get

$$\delta\varepsilon(\tau)_{tot} = \int_{V_{\delta\Sigma}} u(\tau) d^3x + \kappa_{mec}$$

where

$$u(\tau) = \frac{1}{8\pi}\left(E^2 + B^2 - \frac{8\pi}{c}S\right)$$

is the residual local energy density at time $\tau$ in the volume $V_{\delta\Sigma}$.

If the source observed from P were an IS, we should have:

$$S = \frac{c}{4\pi}EB$$

with $E = B$ and consequently

$$u(\tau) = \frac{1}{8\pi}(E-B)^2 = 0 \ . \tag{7}$$

In other words the residual local energy density would be zero, i.e. all the energy produced by the source would be instantaneously emitted. For a RS, instead, $S \equiv S'_r = \frac{c}{4\pi}E_t B$, $S$ being the radial component of the PV and $E_t$ the electric field, tangential to the spherical surface centred at an ideal point-like dipole. For such a source the residual local energy density will be

$$u(\tau) = \frac{1}{8\pi}\left(E^2 + B^2 - 2E_t B\right) \geq 0 \ . \tag{8}$$

Eq. (8) shows that an em source physically comparable with a RS, localises energy in its neighbourhood, i.e. not all the energy produced by the source is instantaneously emitted.

**5. Radial Emission**

The local contribution to the irradiated energy for a RS is given by eq. (3) and it can be usefully analysed (ref. [3]) by introducing the local luminosity vector **Y**, defined as

$$\nabla \cdot \mathbf{Y} = \frac{1}{c}S , \tag{9}$$

so that eq. (3) can be rewritten as

$$\delta\varepsilon_{rad} = \int_{V_{\delta\Sigma}} \nabla \cdot \mathbf{Y} \, d^3x = \oint_{V_{\delta\Sigma}} \mathbf{Y} \cdot d\mathbf{a} .$$

We set formally $|\mathbf{E}\times\mathbf{B}| = R(r)\Theta(\theta)$, where $R(r)$ describes the radial behaviour and $\Theta(\theta)$ the angular one. To get a physically correct behaviour for the energy emission of the source, we must assume

$$R(r) \approx \frac{q^2}{r^4} \ .$$

A different radial dependence from $r$ would yield an unacceptable energy emission.

For a RS the vector $S$ coincides with the radial component $S'_r$ of the PV,

$$S'_r = \frac{cq^2\Theta_r(\theta)}{4\pi \, r^4}$$

where $\Theta_r(\theta)$ is the radial part of the angular distribution.

To analyse only the radial emission of a RS, we rewrite the differential equation (9) in polar coordinates omitting the angular part

$$\frac{1}{r^2}\frac{\partial}{\partial r}r^2 Y(r,\theta) = \frac{e^2 \Theta_r(\theta)}{4\pi r^4}.$$

By setting $z=kr$ and

$$y(z) = \frac{4\pi Y(r,\theta)}{q^2 \Theta_r(\theta) k^3},$$

eq. (9) becomes

$$y_z + \frac{2}{z}y - \frac{1}{z^4} = 0,$$

the solution of which

$$y(z) = \frac{1}{z^3}(\xi_0 z - 1) \tag{10}$$

describes the local luminosity on the spherical surface $\Sigma_{z/k}$.

At great distances (i.e. for $r$ much bigger than the emission wavelength), the emitted wave becomes identical to that emitted by an IS, whereby, in view of the asymptotic behaviour of (10), we obtain that the luminosity on the surface $\Sigma_{z/k}$ for an IS is

$$y_{id}(z) \approx \frac{\xi_0}{z^2}. \tag{11}$$

By comparing eq. (10) with eq. (11), we can see that a RS emits less energy than an IS, the difference of luminosity being

$$\Delta y(z) = y_{id}(z) - y(z) \approx \frac{1}{z^3} \tag{12}$$

This means that an amount of energy proportional to (12) is retained inside the surface $\Sigma_{z/k}$ and localised around the source. Therefore, the residual local energy density $u(\tau)$ for the RS results to be different from zero (see eq. (8)), while for an IS this energy density vanishes (see eq. (7)).

Since the energy emission from the source is continuous, a characteristic balance surface must exist for which the energy emitted across the surface is equal to than not yet emitted. Since the wavelength characterises the period of the wave, we assume that the balance is reached on the wave front for $r = \lambda$. Using eqs. (10) and (12), we can then write the balance condition as

$$\Delta y(k\lambda) = y(k\lambda),$$

the solution of which yields $\xi_0 = 1/\pi$.

The luminosity $Y(r,\theta)$ turns out to be

$$Y(r,\theta) = \frac{q^2}{4\pi r^3}\left(\frac{2r}{\lambda} - 1\right)\Theta_r(\theta). \qquad (13)$$

The r.h.s of eq. (13) becomes zero at $r_0 = \lambda/2$ and reaches its maximum at $r_{max} = 3\lambda/4$. These radii characterise two spherical surfaces $\Sigma_0$ and $\Sigma_{max}$ that delimit a spherical crown $\delta\Sigma$ containing the source zone, i.e. the portion of space where energy is produced. By using the characteristic time of emission $T = \lambda/c$ we define, in terms of time,

$$t_{max} = \frac{2}{c}(r_{max} - r_0) = \frac{T}{2} \qquad (14)$$

which is the threshold time before which the source is active (i.e. it produces energy) and the superficial luminosity increases; after $t_{max}$ the luminosity decreases to the initial em background (see fig.1).

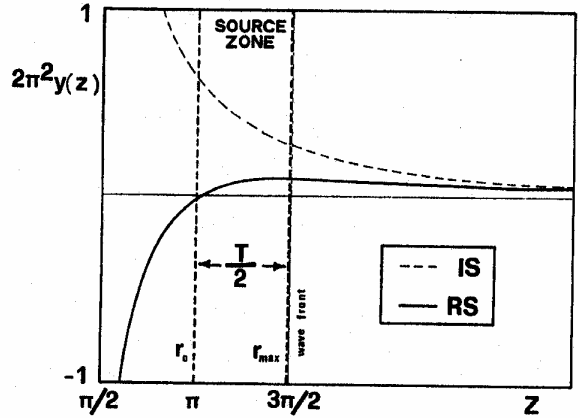

Fig. 1.

## 6. Energy and Momentum

For a RS, the transverse component of the PV is responsible for an anomalous circular propagation of the field on the expanding spherical surface $\Sigma(t)$, that corresponds to the external ideal frontier of the source. The resulting path of the field is described in space by the curve

$$\gamma(t) = \theta(t)r(t).$$

An observer placed on the spherical surface sees the field propagating on $\Sigma(t)$ with speed

$$\dot{\gamma}(t) = \dot{\theta}(t)r(t) = \omega\, r(t) = c. \qquad (15)$$

The speed $\dot{\gamma}(t)$ has been written by deriving only $\theta(t)$, since the observer in his frame of reference sees just a transverse propagation of the em field with speed $c$. The radius $r(t)$ of the expanding surface is the distance of the observer from the centre of the source at time $t$ and it can be written as

$$r(t) = v_{exp} \, t \, , \tag{16}$$

where $v_{exp}$ is the expansion speed of the surface $\Sigma(t)$.

The spatial evolution of this surface defines the source zone, the thickness of which increases from zero up to a maximum given by

$$r_{max} - r_0 = \frac{\lambda}{4}.$$

Using eqs.(15) and (16) we obtain:

$$\dot{\gamma}(t) = \frac{2\pi}{T} v_{exp} \, t = c \, , \tag{17}$$

therefore, when the source luminosity achieves its maximum at $t = t_{max}$ (see fig. 1), the expansion speed is

$$v_{exp} = \frac{c}{\pi}.$$

The source just considered is produced by the interaction of two charges in space. If no other charge is "felt" electromagnetically by the source, we can consider the RS as an isolated system for which the angular momentum

$$l(\theta) = |\mathbf{r} \times \mathbf{p}| = r \, p_t(r,\theta) \tag{18}$$

is constant. Here $p_t$ is the transverse component of the field momentum. Time derivative of eq. (18) must be zero:

$$\dot{l}(\theta) = \dot{r} \, p_t(r,\theta) + r \, \dot{p}_t(r,\theta) = 0 . \tag{19}$$

By defining the transversal momentum on a generic spherical surface $\Sigma_r$ centred on the source as (see ref. [3] and [5])

$$\mathbf{p}_t = \frac{1}{c^2} \int_{V_{\Sigma_r}} \mathbf{S}_t \, d^3x \tag{20a}$$

where

$$\mathbf{S}_t = \frac{cq^2}{4\pi} \Theta_t(\theta) \frac{\hat{u}_t}{r^4},$$

we get:

$$\mathbf{p}_t = -\frac{q^2}{c} \Theta_t(\theta) \frac{\hat{u}_t}{r} \, , \tag{20b}$$

were the expression (20b), for the momentum length $p_t$, satisfies eq. (19) for the conservation of angular momentum. The two terms appearing in eq (19) can be interpreted as two opposite energies, the first due to the expansion of the source zone, the second due to the field spin:

$$-2T_{exp} + 2T_{spin} = 0 . \tag{21}$$

We can see that conservation of the angular momentum requires the simultaneous presence of an expansion of the source zone along with a field spin effect, i.e. the field spin is produced at the expense of the source localisation, since the spatial dimension of the source grows.

As we showed above, when the luminosity achieves its maximum the expansion occurs at a speed $v_{exp} = c/\pi$ smaller than $c$; this allows us to conjecture the existence of a temporary "inertial effect" only during the source expansion, i.e. one may suppose that inside the source zone a hidden "mass" acts gravitationally producing a weak field associated with the source zone of a RS.

The energy associated with the spin effect is given by

$$T_{spin} = |\mathbf{p}_t| c$$

where $c$ is the propagation speed of the field on the expanding spherical surface. From eqs. (21) and (20a,b), integrating by shells over the volume $V_{\delta\Sigma}$ of the spherical crown of the active zone, we get

$$T_{spin} = T_{exp} = \frac{1}{c} \int_{V_{\delta\Sigma}} |\mathbf{S}_t| d^3 x = \frac{q^2}{3\pi} \Theta_t(\theta) k \,, \tag{22}$$

where $k$ is the wave number of the source. Therefore, the local contribution to the spin and to the expansion momentum is given by

$$\delta p_{spin} = \delta p_{exp} = \frac{2T_{exp}}{v_{exp}} = \frac{2q^2}{3c} \Theta_t(\theta) k \,. \tag{23}$$

By integrating once more, over all the contributions associated with the angles of emission, we finally get the total momentum

$$p_{spin} = p_{exp} = \int_{P_\Sigma} \delta p_{exp} = \left( \frac{2}{3} \frac{q^2}{c} \int_0^{2\pi} d\varphi \int_0^\pi \Theta_t(\theta) d\theta \right) k \,.$$

We may identify the constant factor in brackets

$$\mathrm{H}_{spin} = \frac{4\pi}{3} \frac{q^2}{c} \int_0^\pi \Theta_t(\theta) d\theta \tag{24}$$

as the contribution to the total action due to the field. This result shows that the total momentum is proportional to the wave number $k$ through an action constant $\mathrm{H}_{spin}$:

$$\mathbf{p}_{exp} = \mathbf{p}_{spin} = \mathrm{H}_{spin} \mathbf{k} \,.$$

Besides $\mathrm{H}_{spin}$, one has to consider two more contributions to the total action: $\mathrm{H}_{mec}$, due to the Coulomb interaction between the charges, and $\mathrm{H}_{grav}$, produced by the gravitational self-interaction produced by the inertial part of the source. However, since the gravitational term is negligible in this context with respect to $\mathrm{H}_{spin}$ and $\mathrm{H}_{mec}$, the total action can be written as:

$$h = \mathrm{H}_{spin} + \mathrm{H}_{mec} + \mathrm{H}_{grav} \cong \mathrm{H}_{spin} + \mathrm{H}_{mec}. \tag{25}$$

As showed in ref. [4], the global action (25) agrees with the Planck's constant value. However, now we will continue the formal exposition of the theory without taking into account the numerical value of the constant (25).

## 7. Uncertainty Principle for a RS

Let us consider the following null local field quantity:

$$a = \frac{E^2 + B^2 - 2|\mathbf{E} \times \mathbf{B}|}{8\pi} = \frac{E^2 + B^2 - 2|\mathbf{E}_t \times \mathbf{B} + \mathbf{E}_r \times \mathbf{B}|}{8\pi} = 0$$

calculated on the effective non-spherical wave surface. From eq. (8), the quantity $a$ satisfies

$$a \geq u(\tau) - \frac{|\mathbf{E}_r \times \mathbf{B}|}{4\pi},$$

i.e.

$$u(\tau) \leq a + \frac{|\mathbf{E}_r \times \mathbf{B}|}{4\pi} \leq a + \frac{|\mathbf{E} \times \mathbf{B}|}{4\pi}.$$

Therefore, being $a = 0$:

$$u(\tau) \leq \frac{|\mathbf{E}_r \times \mathbf{B}|}{4\pi} \leq \frac{|\mathbf{E} \times \mathbf{B}|}{4\pi},$$

and by integrating over the volume of the spherical crown $V_{\delta\Sigma}$, we get

$$\int_{V_{\delta\Sigma}} u(\tau) d^3x \leq \frac{1}{c} \int_{V_{\delta\Sigma}} S_t\, d^3x \leq \frac{1}{c} \int_{V_{\delta\Sigma}} S\, d^3x. \tag{26}$$

By using eqs. (18) and (20) to describe the field angular momentum, eq. (26) can be written as

$$\delta\mathrm{E}_{res}\tau \leq \mathrm{l}(\theta) \leq \delta\mathrm{E}_{field}\tau,$$

by further integrating over all the angles of emission, at distances $r$ of the order of the wavelength $\lambda$, one gets

$$\mathrm{E}_{res}\,\tau \leq \mathrm{H}_{spin} \leq \mathrm{E}_{field}\,\tau, \tag{27}$$

where the elapsed time $\tau$ is of the same order as the characteristic period of emission T.

According to eq. (25), the total action $h$ contains a contribution $\mathrm{H}_{mec}$ too, given by

$$\mathrm{H}_{mec} = \mathrm{E}_{mec}\,\tau,$$

therefore, eq. (27) yields

$$(\mathrm{E}_{res} + \mathrm{E}_{mec})\tau \leq h \leq (\mathrm{E}_{field} + \mathrm{E}_{mec})\tau. \tag{28}$$

In terms of momentum, the first inequality says that, for $r < \lambda$, the total momentum, measured by an observer inside the source, satisfies the relationship

$$P_{tot}^{(int)} r < h, \qquad (29)$$

while, for an external observer, i.e. for $r \geq \lambda$, the momentum measured is subject to the inequality

$$P_{tot}^{(ext)} r \geq h. \qquad (30)$$

Eq. (29) refers to the momentum measured within the first wave front, i.e. in the zone where the residual energy is localised, and it applies to an observer in presence of a source emitting with a wavelength in the radio zone. For sources emitting in the sub-radio zone, instead, it is not possible to realise an experimental apparatus able to perform observations at a distance shorter than the wavelength. In this case, the momentum satisfies eq. (30), which is similar to Heisenberg's uncertainty principle and expresses the fact that the spherical surface of the first wave front becomes a sort of spatial limit that makes the source look like a compact energy grain with action $h$. We will come back on this subject later on in the paper.

**8. Spatial Bounds in the EM Interaction During a Charge Pair Collision**

When radiation passes through matter, the most frequent mechanisms of energy loss are excitation and ionization of atoms. To calculate the rate of energy loss, one usually makes the following hypotheses: (a) the electrons are free and at rest, (b) the interaction lasts such a short time that the electron gets an impulse without significantly changing its spatial position during the collision. For reasons of symmetry, these assumptions imply that the impulse acquired by the target after collision must be finite and perpendicular to the trajectory of the incident particle and that the interaction must occur within precise spatial bounds. Since the theoretical predictions for energy loss are all in good agreement with experimental observations, the hypotheses (a) and (b) prove to be correct and necessary to explain the energy loss rate. By extending these arguments to include the phenomenology of the source formed by the collision of two electrically charged particles, we shall still assume hypothesis (a) as true and we show that what hypothesis (b) asserts, is already built in the model.

The em emission produced by the pair collision under hypothesis (a) is the simplest physical situation in which a RS can be produced. If we assume that during the collision the impact parameter remains positive, the dipole momentum generated in the interaction is non null and the PV is not everywhere radial. In this case there is a field spin produced by the transverse PV component, the action bounds of which agree with eq. (27). If the elapsed time $\tau$ equals the collision time $t_{col}$ and the action constant is written according to eq. (25), we get for the localised energy in the neighbourhood of the source (see eq. (28)):

$$E_{res} + E_{mec} \leq \frac{h}{t_{col}} \leq E_{field} + E_{mec}. \qquad (31)$$

If we assume $h$ to be finite, the collision time has to be limited and inversely proportional to the particle energy, because if $t_{col}$ were infinite, from eq. (31), $E_{res}$ would be zero, the PV would be radial everywhere, with the effect of yielding an IS. Therefore, the em source produced in this case is a RS, for which $E_{res} > 0$, and not an ideal one.

According to hypothesis (a), we describe the interaction of a charge $q$ with a anti-charge $\bar{q}$ assuming the target particle at rest. If $\mathbf{R}$ is the vector connecting the particles of charge $\bar{q}$ at rest and $q$ moving along the trajectory $r(t)$, the interaction at time $t$ occurs at the conditions of time $t' < t$, usually the charge position $r' = r(t')$ at time $t'$ is delayed by a time $\Delta t = t - t'$ necessary for the signal to propagate along $\mathbf{R'} = \mathbf{R}(t')$ (see fig. 2). Let us call $R' = |\mathbf{R'}|$ the "effective" distance of interaction, and $R = |\mathbf{R}|$ the "actual" distance between the charges. If we arbitrarily assume that the particles achieve the minimal distance $R(0) = \lambda$ when $t=0$, the interaction acts with a delay. Then the "effective" position $R'$ depends both on speed and angle of incidence at a previous time $t' < 0$.

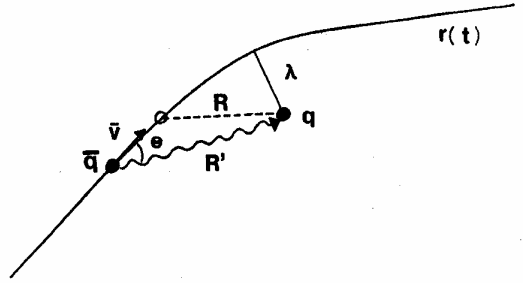

Fig. 2.

We can get an estimate of $R'$ by writing

$$R' \leq \frac{v}{c} R' \cos\theta + R ,$$

that yields

$$R' \approx R\left(1 + \frac{v}{c}\cos\theta\right).$$

Let, the particles achieve the minimal distance $\lambda$ at time $t=0$. The "effective" distance depends on the way the charges approach to each other:

$$R' \approx \lambda\left(1 + \frac{\mathbf{v}\cdot\hat{n}}{c}\right). \tag{32}$$

Denote by

$$t_{delay} = \lambda \frac{\mathbf{v}\cdot\hat{n}}{c^2}$$

the delay time along the direction of the dipole moment varying between 0 and $T = \lambda/c$.

$t_{delay}$ represents the time necessary for the signal to propagate along the dipole axis from the "effective" position $R'$ to the actual position $R$. From eq. (32), we get:

$$R' \approx R(t_{delay}) = \lambda\left(1 + \frac{t_{delay}}{T}\right) . \tag{33}$$

We assume the instant of minimal "effective" distance $\lambda$ between the charges to be the instant of maximum luminosity of the source. From eqs. (32-33), when the velocity of the impinging particle respect to the target is within the interval $0 \leq v < c$, the delay time is always within the interval $0 \leq t_{delay} < T$. Since, from eq. (33), $R(t_{delay}) = \lambda$, when $t_{delay} = 0$, we write

$$1 \leq \frac{R(t_{delay})}{\lambda} < 2 \ .$$

We also know (eq. (15)) that the interval between the instant at which the source starts emitting and the instant of maximum luminosity is equal to $T/2$. Therefore, when $t_{delay} \equiv t_{max} = T/2$, the source starts emitting and the particles are at a distance $3\lambda/2$ apart.

We point out that the bounds are the same as those obtained for the active source zone $\delta\Sigma$ (sect. 5), but if we want to correlate the bounds of the source zone with those of the scattering process described above, we must invert the extremes of the interval. We recall that an increase of the source luminosity occurs during the approach phase between the charges, while the source luminosity decrease during the removal phase. In particular, when the "effective" distance between the charges is $3\lambda/2$, the emission starts with null luminosity from the surface $\Sigma_0$ (with diameter $\lambda$); conversely, when the charges achieve the minimum interaction "effective" distance $\lambda$, the luminosity from the surface $\Sigma_{max}$ (with diameter $3\lambda/2$) is maximum. With these assumptions, we may neglect the interactions occurring outside the source zone, because they do not contribute to the energy production.

For these reasons, we also interpret $t_{delay} = T/2$ as the delay time at which the interaction begins to contribute to the energy production. During the collision, the "effective" distance of interaction varies within the interval

$$R_0 \leq R(t_{delay}) \leq R_{max} \qquad (34)$$

where

$$R_0 = R(0) = \lambda$$

and

$$R_{max} = R(t_{max}) = 3\lambda/2 \ .$$

Therefore, while the source zone of an IS is point-like, the source zone of a RS is a finite spherical crown of diameters $[\lambda, 3\lambda/2]$ referring to the ideal centre. The previous definition of the source zone is a consequence of the fact that the PV varies radially as $r^{-4}$.

Such a peculiar source zone produces a finite interaction time between the moving charges. The time necessary for an incoming charge to move from the actual position along the trajectory, where the distance of interaction is $R_{in} = R_{max}$ before the "turning point", to the symmetrical one $R_{out} = R_{max}$ after the "turning point", will be called "collision time"

$$t_{col} = 2t_{max} = T \ .$$

During the collision time, while the charge moves along the trajectory from the actual distance $R_{in}$ to the "turning point" $R_0$, we have the "build-up" of energy inside the source. In this first phase, the energy inside the source grows, while, from the "turning point" to the actual position of the outcoming charge, the energy stops growing and is just emitted. During this latter phase, the energy is emitted continuously till it reaches asymptotically the way of emission of an IS. According to the theory (sect. 3), there are no contributions to the production of energy, along the trajectory, from the "turning point" (where the actual distance between charges is $R_0$) to the "outcoming point", symmetrical to $R_{in}$ with respect to the "turning point".

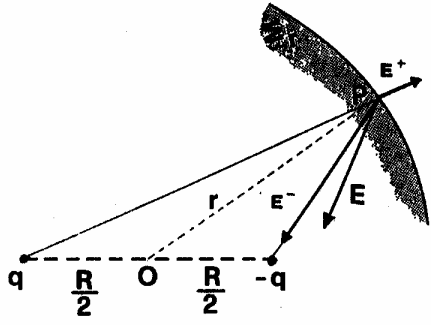

Fig. 3.

## 9. The PV Behaviour in the $q\bar{q}$ Collision

We have seen by examining the electric interaction that occurs during a $q\bar{q}$ charge collision, that the energy contributions to the local source are present just for a limited time $T/2$. Therefore, the em source turns out to be confined within a limited spatial zone bounded by a spherical crown with diameters $\lambda$ and $\lambda + t_{max} c$. The zone of the source energetically active is defined as the source zone.

Let us now consider a spherical surface $\Sigma_r$, of radius $r$, inside the source zone. The local electric field $\mathbf{E}$, measured at a point P (see fig. 3) on the spherical surface, is a superposition of the electric fields $\mathbf{E}^+$, $\mathbf{E}^-$ produced by colliding charges:

$$\mathbf{E} \equiv \mathbf{E}_{(2,1)} = \begin{pmatrix} E^+ \\ E^- \end{pmatrix} = \begin{pmatrix} \dfrac{q}{\left|\mathbf{r} + \dfrac{1}{2}\mathbf{R}\right|^2} \\ \dfrac{-q}{\left|\mathbf{r} - \dfrac{1}{2}\mathbf{R}\right|^2} \end{pmatrix} \quad (35)$$

where $\mathbf{r} \equiv \overrightarrow{OP}$ is the vector joining P to the virtual origin of the source and $\mathbf{R}$ is the dipole moment per unit charge, with length value $|\mathbf{R}| = R(t_{delay})$ (see eq. (33)).

Let us denote by $\rho = \dfrac{R}{r}$ the ratio between $R$ and $r$. Since, according to eq. (34), $R$ varies within the characteristic interval of the distances of interaction, $\rho$ will always be within the corresponding interval

$$\frac{\lambda}{r} \leq \rho \leq \frac{3\lambda}{2r} \quad .$$

Therefore, The electric field of eq. (35) can be written in terms of $\rho$ as

$$\mathbf{E}_{(2,1)} = \frac{q}{r^2} \begin{pmatrix} \eta^+(\rho,\theta) \\ \eta^-(\rho,\theta) \end{pmatrix} \qquad (36)$$

where

$$\eta^\pm(\rho,\theta) = \frac{\pm 4}{4 + \rho^2 \pm 4\rho \cos\theta}$$

are the angular distributions of the electric fields produced by each charge. At a point P of the surface $\Sigma_r$, the electric field (36) can be broken into three components according to an orthogonal local triad $\mathbf{O}_P \equiv (\hat{l},\hat{t},\hat{r})$

$$\mathbf{E} \equiv \mathbf{E}_{(3,1)} = \mathbf{T}_{(3,2)}\mathbf{E}_{(2,1)}$$

using the transformation matrix (see ref. [4])

$$\mathbf{T}_{(3,2)} = \begin{pmatrix} 0 & 0 \\ \dfrac{\rho \sin\theta}{(4+\rho^2+4\rho\cos\theta)^{1/2}} & \dfrac{-\rho \sin\theta}{(4+\rho^2-4\rho\cos\theta)^{1/2}} \\ \dfrac{2+\rho\cos\theta}{(4+\rho^2+4\rho\cos\theta)^{1/2}} & \dfrac{2-\rho\cos\theta}{(4+\rho^2-4\rho\cos\theta)^{1/2}} \end{pmatrix}.$$

The electric field at P is

$$\mathbf{E} = \frac{q}{r^2} \mathbf{\Psi}_{el} \quad, \qquad (37)$$

which is modulated by the vector

$$\mathbf{\Psi}_{el} = \begin{pmatrix} 0 \\ \psi_t \\ \psi_r \end{pmatrix} = \begin{pmatrix} 0 \\ \dfrac{4\rho \sin\theta}{(4+\rho^2+4\rho\cos\theta)^{3/2}} + \dfrac{4\rho \sin\theta}{(4+\rho^2-4\rho\cos\theta)^{3/2}} \\ \dfrac{4(2+\rho\cos\theta)}{(4+\rho^2+4\rho\cos\theta)^{3/2}} - \dfrac{4(2-\rho\cos\theta)}{(4+\rho^2-4\rho\cos\theta)^{3/2}} \end{pmatrix}$$

referred to the local triad $\mathbf{O}_P$.

From eq. (5), we can see that near the source the radial electric field is non-zero. The transverse component of the PV exists and it is responsible for a spin of the em field that has the effect of reducing the radial emission and of localising the non-irradiated energy in the neighbourhood of the source. We could then break the em wave into two different waves: the first, a spherical one, which is responsible for the energy propagation, the second, a spinning wave, which is responsible for the energy localization.

Since the magnetic field in the wave zone is

$$\mathbf{B} = \begin{pmatrix} -E \\ 0 \\ 0 \end{pmatrix},$$

it follows, from eq. (37)

$$\mathbf{B} = \frac{q}{r^2} \mathbf{\Psi}_{mg} = \frac{q}{r^2} \begin{pmatrix} -\Psi_{el} \\ 0 \\ 0 \end{pmatrix}$$

where $|\mathbf{\Psi}_{mg}| \equiv |\mathbf{\Psi}_{el}| = \Psi$ explicitly

$$\Psi = \Psi(\rho,\theta) = 4 \left\{ \frac{1}{(4+\rho^2+4\rho\cos\theta)^2} + \frac{1}{(4+\rho^2-4\rho\cos\theta)^2} - \frac{2(4-\rho^2)}{[(4+\rho^2)^2 - 16\rho^2\cos^2\theta]^{3/2}} \right\}^{1/2}$$

is the modulus of the modulation vectors of the electric and magnetic field of the source. It follows that the local PV is

$$\mathbf{S} = \frac{c}{4\pi} \mathbf{E} \times \mathbf{B} = \frac{q^2 c}{4\pi r^4} \mathbf{f},$$

where the dimensionless vector $\mathbf{f}$, obtained from the product

$$\mathbf{f} = \mathbf{\Psi}_{el} \times \mathbf{\Psi}_{mg} = \begin{pmatrix} 0 \\ f_t \\ f_r \end{pmatrix} = \begin{pmatrix} 0 \\ -\Psi \psi_r \\ \Psi \psi_t \end{pmatrix},$$

is responsible for the polar modulation of the energy flux in the source zone. For this reason, we define

$$\Theta_t(\rho,\theta) = \Psi |\psi_r|$$

as the angular distribution of the energy localised in the source zone. This distribution is associated with the transverse component of the PV that characterises the energy carried around by a spinning wave centred on the virtual origin of the source. The angular distribution of the energy carried out radially from the source, instead, is described by the radial component of the modulation vector responsible for the REM effect, so we write

$$\Theta_r(\rho,\theta) = \Psi |\psi_t|.$$

This angular distribution, modulates polarly the intensity of the energy emission of the source, i.e. the energy that an observer can measure or, in other words, the radial emission of light. Both the distributions depend on the intrinsic structure of the transverse and radial em fields of the source (see fig. 4-5).

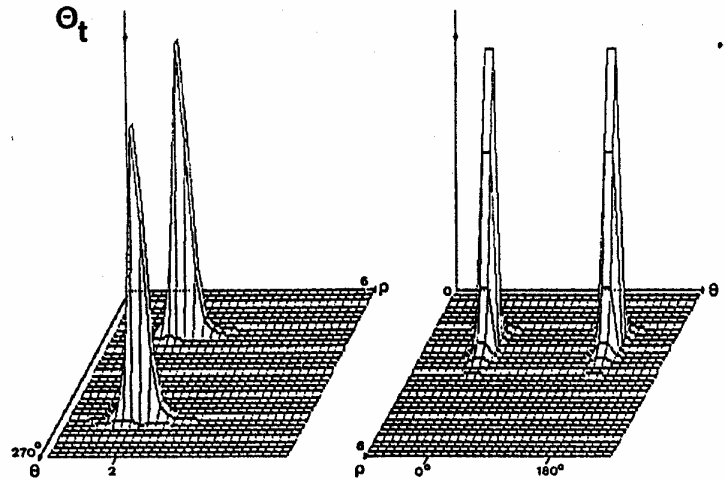

Fig. 4.

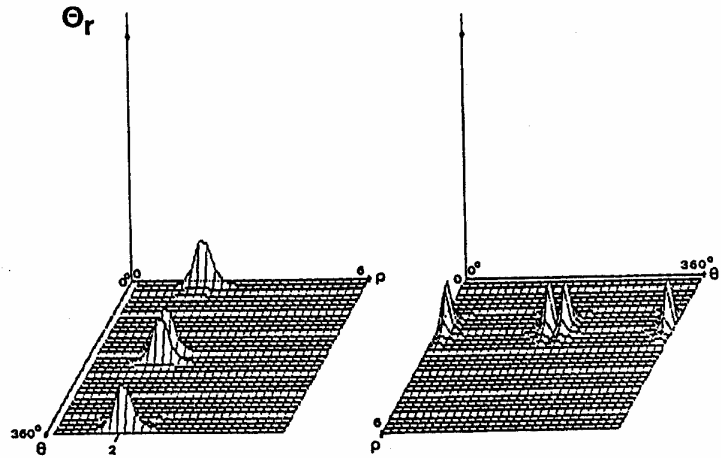

Fig. 5.

Calling respectively $\Theta_t$ and $\Theta_r$ the transverse and radial field structure functions, the components of the PV directly responsible for the amounts of energy localised and emitted, become:

$$S_t = \frac{q^2 c}{4\pi r^4} \Theta_t$$

$$S_r = \frac{q^2 c}{4\pi r^4} \Theta_r .$$

## 10. q$\bar{\text{q}}$ Dynamic

While the particles go through the collision, they make a complete oscillation along the dipole axis with period equal to the collision time $t_{col} = T$ and mean amplitude equal to the dipole moment length $\overline{R} = \langle |d| \rangle / q$, characterising the em emission, the value of which must be determined.

Because of the delay effect in the signal propagation due to the finite speed of light, the charges at time *t* are subject to the source field produced at a previous time *t'* when the dipole moment of the source was **d**'. In other words, the source "feels" the past state of the em field and one can assume that the interaction occurs as if the field of the source were rotated by an angle roughly equal to the mean angle of incidence (or diffusion). We visualise this effect as a torsion of the em field of the source.

We make now two observations: from the geometry of the process, the values of the angles of incidence and diffusion are always within the interval $[0, \pi/2]$, the non-point-like charge spatial distribution of the incident charge *q* determines an angular spread of charge in the region in which the retarded field begins to act. The angular zone in which the charge is spread extends reasonably for just a few seconds of arc. Therefore, the interaction occurs in the range of eq. (34), with an angle of incidence greater than the mean angle $\bar{\theta} = \pi/4$ by some seconds of arc. The mean angular value, as an effect of the charge spread, requires therefore a very little, but non easily valuable correction. In this sense the mean angle should be considered as a lower angular bound.

The dipole in its "actual" position is subject to a Lorentz's force **F** produced by the em field of the source in its "effective" position (see fig. 6). Since for the source $|\mathbf{E}| = |\mathbf{B}| \equiv E$, the Lorentz's force value is

$$|\mathbf{F}| = q|\mathbf{E} + \boldsymbol{\beta} \times \mathbf{B}| = qE|\hat{e} + \beta\,\hat{v} \times \hat{b}|$$
$$= qE|\hat{e} + \beta\hat{R}| = qE(1 + \beta^2 + 2\beta\cos\vartheta_{eR})^{1/2} \;,$$

that we can write in general $|\mathbf{F}| = qE\zeta(\beta, \vartheta_{eR})$. The resulting force component, acting on the moving charge along the direction of the dipole axis, has the effect of varying the strain on the

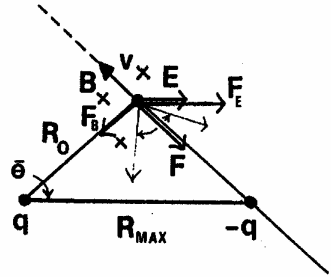

Fig. 6.

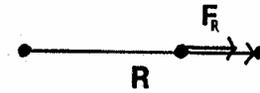

Fig. 7.

dipole. The acting force is $F_R = \mathbf{F} \cdot \hat{R}$ (see fig. 7). Since

$$|\mathbf{S}| = \frac{c}{4\pi}|\mathbf{E} \times \mathbf{B}| = \frac{c}{4\pi}E^2 \;,$$

the square mean value of this force during the collision time is

$$\left\langle F_R^2 \right\rangle = \frac{\int\limits_{\lambda}^{3\lambda/2} F_R^2 \, dR}{\int\limits_{\lambda}^{3\lambda/2} dR} = q^2 \zeta(\beta, \vartheta_{eR})^2 \cos^2 \vartheta_{FR} \frac{\int\limits_{\lambda}^{3\lambda/2} E^2 \, dR}{\int\limits_{\lambda}^{3\lambda/2} dR}$$

$$= \frac{4\pi q^2}{c} \zeta(\beta, \vartheta_{eR})^2 \cos^2 \vartheta_{FR} \frac{\int\limits_{\lambda}^{3\lambda/2} |\mathbf{S}| \, dR}{\int\limits_{\lambda}^{3\lambda/2} dR} .$$

Whereas the square mean value of the work of the field on the moving charges is, analogously,

$$\left\langle W^2 \right\rangle = \frac{\int\limits_{\lambda}^{3\lambda/2} (F_R R)^2 \, dR}{\int\limits_{\lambda}^{3\lambda/2} dR} = \frac{4\pi q^2}{c} \zeta(\beta, \vartheta_{eR})^2 \cos^2 \vartheta_{FR} \frac{\int\limits_{\lambda}^{3\lambda/2} |\mathbf{S}| R^2 \, dR}{\int\limits_{\lambda}^{3\lambda/2} dR} ,$$

We define the dipole mean square length $\overline{R} \equiv \sqrt{\left\langle R^2 \right\rangle}$, as

$$\left\langle \mathrm{R}^2 \right\rangle = \frac{\left\langle W^2 \right\rangle}{\left\langle R^2 \right\rangle} = \frac{\int\limits_{\lambda}^{3\lambda/2} |\mathbf{S}| R^2 \, dR}{\int\limits_{\lambda}^{3\lambda/2} |\mathbf{S}| \, dR}$$

(38)

$$= \left[ \frac{\int\limits_{\lambda/r}^{3\lambda/2r} \Psi(\rho, \overline{\theta})^2 \rho^2 \, d\rho}{\int\limits_{\lambda/r}^{3\lambda/2r} \Psi(\rho, \overline{\theta})^2 \, d\rho} \right] r^2 = \overline{\rho}^2 r^2 .$$

The mean square length represents the "characteristic" distance between the interacting charges during the building up of the source zone; hence $\overline{R}$ is the "characteristic" dipole length of the source.

**11. Dipole Moment**

The spherical wave front emitted from the source is assumed to be the surface on which the balance condition between the emitted and the retained energy is realised (see sect. 4).

Setting $r = \lambda$, the distance of the first wave front from the virtual origin of the source, and using eq. (38), we define $\overline{d} = q\overline{\rho}\lambda$ as the dipole moment mean square length. The value of the parameter $\overline{\rho}$ can be estimated assuming $\overline{\theta} = \pi/4$ as the lower angular limit of the mean torsion of the source em field. Solving, by Clenshaw-Curtis integration, the integrals in square brackets of eq. (38) we get

$$\bar{\rho} = \frac{\bar{d}}{q\lambda} \equiv \frac{\bar{R}}{\lambda} = 1.2755578749164 \ .$$

As an effect of the angular spread of the spatial charge distribution, associated to the particle interacting with the em field of the source, we assume, and will later determine, an inaccuracy in the dipole moment estimate.

Actually, because of the inaccurate knowledge of the amplitude of angular spread, we assume that the charge collision is an em process in which a local RS with the effective mean dipole moment $\bar{d}' \leq q\bar{\rho}\lambda$ and oscillation period $T = \lambda/c$ is created. The effective mean dipole length value and the associated inaccuracy that will be assumed (see tab. 1), are both considered as essentially produced by the effective torsion of the em field due to the delay effect and by the magnitude of the angular spread of a charge seen by the source field subject to the torsion.

Table. 1: Angular variation in column 3 corresponds to an uncertainty on the electromagnetic coupling constant of 0.045 ppm.

| Mean Angle (rad) | Angular Correction (rad) | Angular Variation (rad) | $\bar{\rho}$ | $\alpha^{-1}$ |
|---|---|---|---|---|
| $\pi/4$ | – | – | 1.275557874 | 137.036669 |
| $\pi/4$ | $+2.19 \ 10^{-5}$ | $-2 \ 10^{-7}$ | 1.275556698 | 137.035995 |
| $\pi/4$ | $+2.19 \ 10^{-5}$ | – | 1.275556687 | 137.035989 |
| $\pi/4$ | $+2.19 \ 10^{-5}$ | $+2 \ 10^{-7}$ | 1.275556676 | 137.035983 |

## 12. Nature of the Sommerfeld's and Planck's Constants.

In his book on QED [1], R.P. Feynman writes that a good electrodynamics must lead to estimating from first principles the value of the electromagnetic coupling constant $\alpha$, namely give in a self-consistent way the numerical and physical dependence of its value on the internal parameters of the theory.

A detailed theoretical analysis of the em field structure near the source zone of the RS, allows us to determine the physical dependence of the coupling constant $\alpha$ on the em interaction parameters. Besides, we can correlate the origin of the Planck's constant with one of the coupling constants, i.e. we can prove that the action constant $\hbar$ needs not to be assumed as fundamental. Indeed both the involved constants ($\alpha$ and $\hbar$), as an effect of their em nature, show two internal degrees of freedom associated the first with the dipole moment length of the interacting pair, the second with the variation of the em field torsion produced by the delay effect during pair interaction.

These dependencies show how both the constants cannot be considered absolutely and universally as such, because it may happen that, in a different physical context with respect to that in which the RS is usually produced, they acquire different values.

According to sect. 4, the explicit expression for the momentum produced by the source zone expansion, necessary to keep constant the angular momentum associated with the transverse component of the PV, is given by

$$\mathbf{p}_{exp} = \mathrm{H}_{spin}\,\mathbf{k}.$$

According to eq. (24), $\mathrm{H}_{spin}$ can be written as

$$\mathrm{H}_{spin} = \frac{4\pi}{3}\frac{q^2}{c}F_t \tag{39}$$

where we have defined

$$F_t = \int_0^\pi \Theta_t(\bar\rho,\bar\theta)\,d\theta.$$

The value of this factor results to be
$$F_t = 32.7034188915657\quad.$$

By substituting this value into the expression for the expansion momentum of the source and using eq. (39), we get

$$\mathbf{p}_{exp} = \frac{4\pi}{3}F_t\frac{q^2}{c}\mathbf{k} = 136.987760716016\,\frac{q^2}{c}\mathbf{k}.$$

In terms of energy, this gives

$$\mathsf{E}_{exp} = 136.987760716016\,\frac{q^2}{c}\omega.$$

The mechanical energy produced during the collision increases when the charges move towards the source virtual origin and the energy depends on the electric force acting on the charges along $\mathbf{R}$. In terms of mean square length, the mechanical energy is

$$\mathsf{E}_{mec} = \int_{R_{max}}^{R_0}\mathbf{F}\cdot d\mathbf{R} = -\frac{q^2}{\langle R^2\rangle}\int_{R_{max}}^{R_0} dR$$

$$= \frac{1}{4\pi\bar\rho^2}\frac{q^2}{c}\omega = 0.048909114430\,\frac{q^2}{c}\omega$$

Therefore, the total energy produced during the collision is

$$\mathsf{E}_{source} = \mathsf{E}_{exp} + \mathsf{E}_{mec} = \left(\frac{4\pi}{3}F_t + \frac{1}{4\pi\bar\rho^2}\right)\frac{q^2}{c}\omega$$

$$= 137.036669830447\,\frac{q^2}{c}\omega \tag{40}$$

By using now the expression for the total action constant of eq.(25), we write

$$h \cong H_{spin} + H_{mec} = \left(\frac{4\pi}{3}F_t + \frac{1}{4\pi\bar{\rho}^2}\right)\frac{q^2}{c} = A^{-1}\frac{q^2}{c}, \quad (41)$$

where the constant

$$A = \frac{q^2}{hc} = \left(\frac{4\pi}{3}F_t + \frac{1}{4\pi\bar{\rho}^2}\right)^{-1} = 1/137.036669830447$$

is independent on the values of the electric charge $q$ and of the speed of light $c$ and plays the same role as the fine structure constant.

We point out that $A^{-1}$ depends mainly, through the factor $F_t$, on the geometric structure of the em field produced in space during the pair collision, and only weakly (0.35%) on the electric force produced by the charges.

In standard QM, the action $\hbar$, the value of which is measured only experimentally, is assumed as fundamental, so that $\alpha^{-1}$ becomes a function both of the electron charge and of the speed of light. Instead, according to eq. (41), it is the constant $A^{-1}$, which has the same role as $\alpha^{-1}$, that results to be fundamental. By this we mean that in ordinary conditions its average value can be assumed as a constant not depending on $q$ and $c$, while $h$, that play the same role of $\hbar$, depends on the values of $A^{-1}$, $q$ and $c$.

Using definition (41), the total localised energy and momentum exchanged in the collision when the source is produced, become

$$E_{source} = h\omega \; ; \; \mathbf{p}_{source} = h\mathbf{k}. \quad (42)$$

Eqs. (42) are the usual expressions for energy and momentum of a photon. Even though the masses associated with the interacting charges do not appear explicitly, they nevertheless exist, hidden inside the source as localised energy.

The value of the characteristic action constant $h$, as shown in the table below, is in good agreement with the experimental one.

*Experimental value :*

$\hbar = 1.05457266(63) 10^{-34}$ J s  (0.60 ppm).

*Theoretical value :*

$h = 1.05457\mathbf{790206204}\, 10^{-34}$ J s  (0.60 ppm).

In order to get the numerical estimate above, we used for $q$ the electron charge (see ref. [6]), which is responsible for the uncertainty of the theoretical value of the action constant.

## 13. Localization and Angular Spread of an Interacting Charge

Assuming as true the experimental value of the coupling constant (see ref. [6]): $A = \alpha$, we may estimate numerically a correction for the value of the interaction mean angle $\bar{\theta}$.

This value is essential to describe more accurately the physical process taking place during the $q\bar{q}$ interaction.

The numerical estimate of the correct mean angular value gives

$$\bar{\theta}' = \bar{\theta} + \theta_{cor} = 0.250006974535\,\pi \quad rad$$

corresponding to a small correction

$$\theta_{cor} = 2.1911147918\,10^{-5}\,rad \ .$$

Using $\bar{\theta}'$, the expression for the dipole mean square length (see eq. (38)) allows us to evaluate with more precision the values of the constants characterising the average energy and momentum:

$$\begin{aligned}\bar{d} &= 1.275556687241109\,e\lambda \\ A &= 1/137.035989561000 \\ h &= 1.0545726669879110^{-34} \quad J\,s\end{aligned} \qquad (43)$$

Assuming that the uncertainty of the experimental estimate of the coupling constant, 0.045 ppm (see ref. [6]), is due to the effect of angular spread of the interacting charge, we have numerically calculated the upper limit of the mean angular spread of the charge interacting with the em field of the RS. This evidences how the interacting charge appears distributed inside a very thin angular region with extremes

$$\bar{\theta}' \cong \bar{\theta} + \theta_{cor} \pm 2\,10^{-7} \quad rad$$

corresponding to an arc width $\Delta_q = \tfrac{1}{2}\bar{R}\Delta\bar{\theta}' \cong 2.5\,10^{-7}\lambda$ that may be interpreted as the em dimension of the interacting charge particle.

In view of what we have just discussed and from the results obtained for the action value characterising an em RS, we write, within the uncertainty interval (see tab. 1), $h \equiv \hbar$ assuming implicitly, that the action constant $h$ is both formally and conceptually Planck's constant. This assumption allows us to build now, a conceptual and phenomenological bridge between the Classic em Theory and QM.

**14. Planck's Action Variability**

In a RS the effective wavelength of emission varies in time with the dynamic of the interaction between charges (see eq. (32)). In particular, since the maximum of emission of the source is achieved when the charges interact at minimum distance, i.e. when the delay time in the signal propagation is zero, $\lambda$ is equal to the minimum distance achieved by the charges during the RS life.

The mean square distance between the interacting charges is, for a free interaction, dependent on $\lambda$ and $\bar{\rho}$, as shown in the eq.(38) and discussed in sect. 11. The specific dependence of dynamical variables of the interaction on these two parameters, allows us to substitute the time evolution of the RS with its average behaviour during collision time.

Energy and momentum of a RS have been calculated and their expressions agree with QM predictions for a photon, so the precision depends only on the calculation of Planck's constant.

BT proves that the value of the action constant depends on the structure of the em field produced near the RS. When the calculation is performed, the precision achieved is high, but inevitably conditioned by the impossibility of knowing exactly the temporal evolution of the source. The dependence of Planck's constant on the effective spatial evolution of the dipole is important only for an observation time lower than the collision time and negligible for an observation time larger than or equal T, i.e. for an ordinary observation time. When a RS is produced in presence of intense external fields, the ensuing constraints on the dipole moment, impose an anomalous time evolution in the relative motion of charges. In this case, the mean interaction distance is modified with respect to the "free" case. Consequently, the characteristic value of the coupling constant changes, varying the local value of Planck's constant. The local variation of the dipole moment produced by external conditions has the effect of modifing, in eq. (41), the characteristic mean value $\bar{\rho}$. So, by rewriting the equation in terms of $\bar{\rho} = \bar{R}/\lambda$, we may actually rewrite the Planck's action as a local variable and not as a constant. For a precise value of the wavelength the action results to be dependent only on the variation of the maximum value of the interaction distance $R$, but it is more convenient to write Planck's action as:

$$\hbar(\bar{\rho}) = \left( \frac{4\pi}{3} F_t(\bar{\rho}) + \frac{1}{4\pi\bar{\rho}^2} \right) \frac{e^2}{c} \quad . \tag{44}$$

In tab. 2 we show graphically the behaviour of the action (44) in a limited range of values of ρ.

Variations of the trajectory of the impinging charged particles, imposed by external fields, have the effect of modifying the coupling value between charge and em field only near the RS. Such variations do not lead to violations of space-time symmetry, because they are produced in different local conditions, i.e. when the local symmetry is not the same existing in free space.

The local variation of the coupling "constant" is likely to be restricted within a limited range of possible values of ρ. In fact, during a pair collision in presence of external fields, the RS can be produced with different values of $\bar{\rho}$ always within the interval of extremes [1, 2[ : $\bar{\rho} \geq 1$ when the delay time of interaction is close to zero, $\bar{\rho} < 2$ when $t_{delay}$ is positive but smaller than T, i.e. when the velocity v of the impinging particle is such that v<$c$ .

Table.2: The "constant" effect characteristic of a restricted range of numerical values of ρ. This effect for $\alpha^{-1}$ is shown within progressively smaller numerical ranges of the ratio ρ. Significant variations of the coupling value are obtained only for consistent variations of the dipole moment.

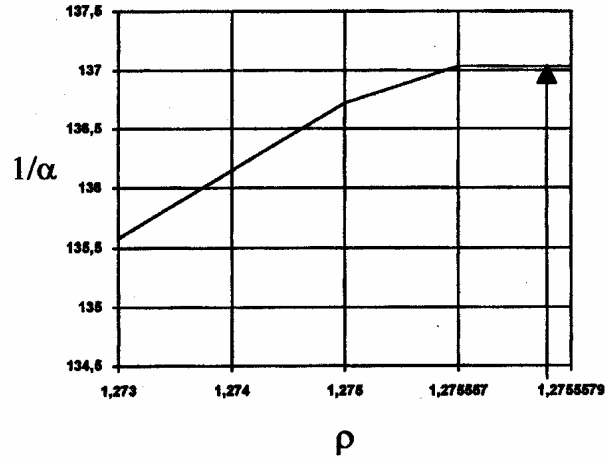

Due to energy and momentum conservation laws, the effective action measured in these cases cannot have values larger than the "free" one, so when the RS is produced, its action acquires presumably a value only a bit smaller than the "free" one. The missing part of action is used by the external fields to act on the interacting charges. There follows that any experimental measure of $\alpha^{-1}$ and $\hbar$ requires an interaction with an external experimental apparatus, so different measures may give different results depending on the experimental method used.

### 15. Classical EM Theory and QM

In the previous sections, we proved the equivalence between $h$ and Planck's constant. Such equivalence puts in evidence how the predictions for the energy and momentum exchange during the em interaction between two massless charges of opposite sign (charge and anti-charge) are able to explain the quantum phenomenology. Moreover, it is our intention to put in evidence how the BT, based on classical EM theory, can physically justify the fundamental QM postulates.

An EM field can always be described as produced by one or more local charges suitably distributed in space. Hence, the interaction between a moving charge and an em field can be described as the interaction of many pairs, each one formed by the moving charge and one of the charges producing the local em field. Each pair is a RS, therefore the energy and the momentum exchanged during the charge-field interaction is the sum of the energies and momenta localised in the neighbourhood of each source.

Following this idea, we now consider a black-body at temperature T formed by a cavity with internal diameter $D$. The em field characterising thermal radiation can exchange energy only with the most external electrons of the atoms placed on the internal surface of the cavity. The field-electron interaction produces a very large number of RS's with a characteristic wavelength within the interval $I_\lambda=]0,D]$. The energy and momentum localised in each RS are a part of the total energy and momentum exchanged between field and matter, i.e. each amount of localised energy and momentum can be thought of as a "photon" with a wavelength value within $I_\lambda$. In this sense the quanta hypothesis, introduced by M. Planck to explain the black-body spectrum, continues to be true, not as a postulate but rather as a physically justified effect in the BT.

The black-body spectrum resulting from the BT, is substantially equivalent to the one predicted by Planck with the exception of a cut-off on the emission wavelength, the value of which must be lower than or equal to $D$. If we consider macroscopic black-bodies, from the experimental point of view, the difference between the two predictions is irrelevant. From the microscopic point of view, however, their difference can become relevant.

An other critical issue deserving attention is the photoelectric effect. The BT predicts that the exchange of energy and momentum between the field of the em wave impinging on a metal surface and the electrons of the conduction band occurs through RS's produced during the direct interaction between the target electrons and a number of anti-charges needed to describe the local intensity of the wave em field . The resulting physical effect is the exchange of amounts of energy and momentum proportional to the frequency of the wave field, i.e. proportional to the inverse of the wavelength of the RS's produced during em collision. A photoelectron is emitted only when the energy received from the em field trough the RS is larger than the characteristic threshold energy for the material, i.e. when the light impinging on the target electrons has a frequency higher than the threshold frequency. According to BT, the total number of photoelectrons emitted is proportional to the number of RS's, i.e. proportional to the number of anti-charges that the em field needs for its local description. In this way the picture describing the photo-emission, agrees both with the photoelectric phenomenology and with QM predictions.

Further aspects characterising QM, such as wave-particle duality or inertial mass, might also be understood by using the "RS method", though a complete theoretical description of these phenomena needs further work, that is now in progress.

In general, the use of the RS's for the description of the em phenomena gives results not in contrast with both classical and quantum phenomenology. This aspect is fundamental for the BT, because it allows to justify the foundations of QM by describing the em interactions as a whole. This shows how the classical and quantum descriptions are intimately connected. Without the BT, each of them is able to describe only one aspect of the physical reality. In fact, Classical em Theory describes macroscopically the space-time evolution of fields induced in single or collective interactions among charges, whereas QM describes microscopically the evolution in terms of energy and momentum of a single interaction between charge and field.

Using the BT, the description of the interactions is relativistically invariant, because it is made through a wave field produced in the RS. Since individual interactions occur between charged massless particles, "inertia" and consequently gravity, would result as an energy localisation in the neighborhood of the charge or of the source, induced by the lack of spherical symmetry in the wave emission, as seen by an observer in relative motion with respect to the RS. In this sense, rest mass would justify the physical inability of the observer to approach indefinitely the source, namely, to measure a localised static electric field.

## 16. The Uncertainty Principle for the RS and the Interacting Particles

In sect. 6, we showed that for any observer able to perform experimental measures on the source emission in the sub-radio zone, the following inequalities are true

$$\mathsf{E}_{tot}\tau \geq \hbar \quad \text{for} \quad \tau \geq T , \tag{45a}$$

$$P_{tot}r \geq \hbar \quad \text{for} \quad r \geq \lambda . \tag{45b}$$

The quantities $\mathsf{E}_{tot}$ and $P_{tot}$ are, respectively, the energy and momentum measured by the observer in the zone external to the RS. They are functions, respectively, of the observation time $\tau$ and of the distance $r$ from the virtual origin of the source. The values of actions $\mathsf{E}_{tot}\tau$ and $P_{tot}r$ may reach the minimum value $\hbar$ only when the observation time $\tau$ is exactly equal to the collision time, $\tau = T$ and, on the same ground, when the distance between source and observer is equal to wavelength, $r = \lambda$, i.e. when the observer is placed on the starting wave front of the source. In view of the causality principle, these conditions are physically unrealizable, because it is experimentally impossible to perform a measuremant while a source is being produced and in the exact spatial position of the first wave front, whose wavelength is a priori unknown.

Hence, eqs. (45) represent a physical limit for any observer. In other words, assuming the minimum action $\hbar$ as an unreachable physical value, eqs. (45) constitute a formulation of Heisenberg's uncertainty principle. Let $\Delta \mathsf{E} = \mathsf{E}_{tot}$ be the total energy variation from the background measured by an observer imbedded in the em external field during the time interval $\Delta t = \tau \geq T$. The total action satisfies then the uncertainty relation

$$\Delta \mathsf{E}\, \Delta t \geq \hbar\,; \tag{46}$$

or, in terms of momentum and position:

$$\Delta P\, \Delta x \geq \hbar\,. \tag{47}$$

Since the actions (46), (47) achieve the minimum value only ideally, all the experiments involving coupled and simultaneous measures of energy and time or momentum and position, are subject to an uncertainty law equivalent to Heisenberg's principle.

### 17. The Spin Angular Momentum

We define angular momentum of the local fields associated to the transverse component of the linear momentum on the first wave front of the RS as

$$\mathbf{l} = \lambda \hat{r} \times \mathbf{p}_t = \frac{2e^2}{3c} f_t(\theta),$$

$\mathbf{l}$ is zero when $\theta$ has the values $\pm \pi/2,\, 3\pi/2$ (see fig. 8).

These angular values define the bounds of the two zones where the interacting charges are located. By integrating over the appropriate angular intervals we get the contribution of the angular momentum spatially associated with each charge.

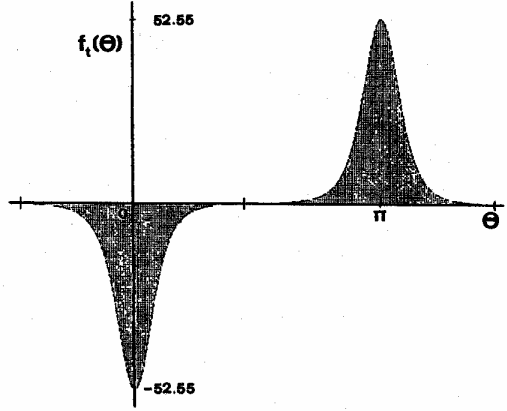

Fig. 8.

Let *L* be the modulus of the total angular momentum associated with each of the two zones occupied by the positive and negative charge, fig. 8. Accordingly, we define

$$L = \int_0^\pi d\varphi \int_{-\pi/2}^{\pi/2} |\mathbf{l}|\, d\theta \quad . \tag{48}$$

Since the action constant depends on the field spin (eq. (39)), we have

$$L = \frac{1}{2} H_{spin} \quad . \tag{49}$$

This equation agrees with the expression for the spin angular momentum of a spin-1/2 particle, except for the value of the action constant. This needs some explanation. If we refer to eq. (41), the action constant consists of two different terms. Eq. (49), instead, depends only on a single action term. In fact, eq. (48) gets contribution just from the field spin, so that the action constant involved in definition (49) is not equal to the free value of the Planck's constant used to define the energy and momentum of a photon (eq. 41) and the spin for the interacting charges cannot be $\frac{1}{2}\hbar$, as QM predictions want.

The difference between the two constants, however is very small and their ratio is very close to one:

$$\frac{H_{spin}}{\hbar} = \frac{136.9873}{137.0362} \cong 0.99964 \ .$$

On the other hand, we should remember that any experiment allowing to measure the Planck's constant is based on energy or momentum measurements, which always give an experimental value for $\hbar$ and not only for $H_{spin}$. Hence the action constant of eq. (49), involved in the definition of the intrinsic angular momentum of the interacting charges producing the RS, cannot be directly measured, because the process of measurement always involves interaction with an external apparatus, with the effect of forcing us to measure $\hbar$ and not just $H_{spin}$.

The angular momenta associated with the zones of space containing the positive and negative interacting charges (IC), in units of $H_{spin}$ are defined as:

$$\mathbf{L}_{IC} = \begin{pmatrix} L^- \\ L^+ \end{pmatrix} = \frac{1}{H_{spin}} \int_0^\pi d\varphi \int_{-\pi/2}^{\pi/2} \frac{2}{3} \frac{e^2}{c} \begin{pmatrix} f_t(\theta) \\ \tilde{f}_t(\theta) \end{pmatrix} d\theta = \begin{pmatrix} -\frac{1}{2} \\ +\frac{1}{2} \end{pmatrix}$$

where $f_t(\theta)$ is the transverse component of the vector $f$ responsible for the polar modulation and

$$\tilde{f}_t(\theta) = -f_t(\theta)$$

is the same component, but for switched charges. In this sense the spin depends on the frame in which the interaction is observed.

Extending the calculation to the complete source zone (SZ) and assuming the dipole axis as axis of symmetry, we have to integrate the angular functions over all the directions, obtaining a null total spin both for unswitched and switched charges:

$$\mathbf{L}_{SZ} \equiv \mathbf{L}_{ph}^{(0)} = \frac{1}{H_{spin}} \int_0^\pi d\varphi \int_{-\pi/2}^{3\pi/2} \frac{2}{3} \frac{e^2}{c} \begin{pmatrix} f_t(\theta) \\ \tilde{f}_t(\theta) \end{pmatrix} d\theta = \begin{pmatrix} 0 \\ 0 \end{pmatrix}.$$

This frame invariance forces us to assume the null spin values of the source as a unique effective component.

Considering em emission of the source, the most probable directions of propagation of the photons are along the wave number $k$ direction that is normal to the dipole axis. Then, for an observer, the angular momentum can be naturally calculated using the propagation axis as axis of symmetry around which the dipole moment is spinning during the interaction. By calling $\varphi'$ the angle measured around this axis, we get

$$\mathbf{L}_{ph}^{(1)} = \frac{1}{H_{spin}} \int_0^{2\pi} d\varphi' \int_{-\pi/2}^{\pi/2} \frac{2}{3} \frac{e^2}{c} \begin{pmatrix} f_t(\theta) \\ \tilde{f}_t(\theta) \end{pmatrix} d\theta = \begin{pmatrix} -1 \\ +1 \end{pmatrix}.$$

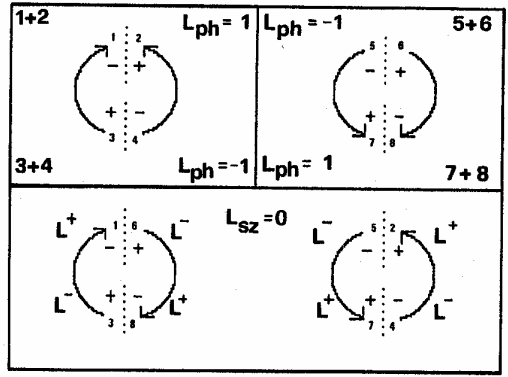

Fig. 9.

The two components of this vector are the spin components corresponding respectively to the left and right circular polarization of the emitted em wave, i.e of the emitted real photons.

During pair interaction, we can put in evidence two sequential phases: (a) the RS is produced, i.e. an amount of energy and momentum of the interacting pair is localised inside the source zone, (b) the source stops to exist by emitting a charge-pair or a photon-pair. In both phases the conservation of angular momentum requires that the spin must be obtained by using the components of the vector $\mathbf{L}_{IC}$ (see fig. 9). Therefore, we can formally write the spin associated to the two phases as

$$L_{ph}^{(0)} = L^- + L^+ = 0,$$

for phase (a) and

$$L_{ph}^{(-)} = L^- + L^- = -1$$

$$L_{ph}^{(+)} = L^+ + L^+ = +1$$

for phase (b).

On the other hand from the point of view of symmetry, there exist two natural axes directly involved in the physical evolution of each phase. The former is the dipole axis, which is the proper axis of symmetry of the source, the second is the emission axis, which gets involved when two photons are emitted. We argue that the value of angular momentum associated to phase (a) "photon exchage" or phase (b) "photon emission", has to do with these axes.

Therefore, by summarising the spin behaviour in a unique vector, we write

$$\mathbf{L}_{ph} = \begin{pmatrix} L^- + L^- \\ L^- + L^+ \\ L^+ + L^+ \end{pmatrix} = \begin{pmatrix} -1 \\ 0 \\ +1 \end{pmatrix},$$

where the null component of the angular momentum allows us to look at the SZ as a massive intermediate boson-state realised during the charge interaction, while the other non null components of the vector $\mathbf{L}_{ph}$ could be assumed to be the angular momenta of the photons associated with em wave emission occurring when the intermediate state ends.

**18. What is the Classical Limit**

The BT encompasses in itself both quantum and classical aspects. In fact, its theoretical and conceptual structure is compatible with Classical em Theory, but also with QM. From the macroscopic point of view, the phenomenology involved in the BT results to be the same as that described by classical theory. From the microscopic (local) point of view, the BT, by reducing the level of approximation of the theoretical description of the em emission, becomes also compatible with quantum theory. This compatibility, coming from the quantum-like way of exchanging energy and momentum during the em interactions between charges and field, vanishes only when we neglect the transverse component of the PV associated to each RS. Such an approximation is equivalent to the classical limit in QM.

Because of these peculiar aspects, we can consider the BT as a further confirmation of the validity of the Classical em Theory, but also as a natural support for a revised QM.

**19. Conclusions**

The results of this work lead us to two orders of general considerations: first, about the physical nature of "constants", second, about the bases of Classical electromagnetism and of QM.

Let us start considering the constants. Is it possible that the fundamental constants, considered to be absolutely the same in the entire universe, are not really constant quantities? This may in fact be the case, because we think that complex physical phenomena able to account for the value of each constant could be found and that the dependence on hidden phenomena could be ascertained for all the "physical constants", like we did, in the framework of the BT, for both Sommerfeld's and Planck's constants. There, the combined effects of natural restrictions for the range of the numerical variation and the instrumental limits characterising the experimental measuring methods, that operate a numerical cut-off, force slowly varying physical quantities to appear to the observer as "constants".

As for the physical bases of Classical em Theory and QM. Since the experimental value of the em "fundamental constants" depends on the values of critical variables, the study of the phenomenology responsible for the existence of "pseudo-constants" represents the access-door to a deeper understanding of the real nature of phenomena. In this sense, the BT describes the "quantum-like nature" of the microscopic em effects, as QM does, and cannot be considered a tautology on any ground. Moreover, the study of its phenomenology could become the key to a deeper comprehension of the role that Classical em Theory and QM play to describe the macroscopic and microscopic aspects of electromagnetism. Namely, the BT could allow a theoretical unification of all em phenomena, overcoming the conceptual difficulties introduced by the "ad hoc" quantum Planck's hypothesis, foreign to classical electromagnetism, but essential to make the theoretical description of phenomena consistent with physical reality.

**Acknowledgement**

The authors wish to thank Prof. G. Bologna of the University of Torino for encouragement and for the helpful discussion of very particular aspects of this non-standard theory.